\documentclass[%
 twocolumn, reprint, showpacs,
 amsmath,amssymb,
 aps, pre, superscriptaddress, floatfix
]{revtex4-1}

\usepackage{graphicx}
\usepackage{dcolumn}
\usepackage{bm}
\usepackage{amsmath}
\usepackage{amssymb}
\usepackage{graphicx} 
\usepackage{chemarr}
\usepackage[caption=false]{subfig}

\newcommand{\circled}[1]{\protect\raisebox{.5pt}{\textcircled{\protect\raisebox{-1pt} {#1}}}}

\begin{document}

\title{Tethered single-legged molecular spiders on independent 1D tracks}

\author{David Arredondo}
\affiliation{Nanoscience and Microsystems Engineering, University of New Mexico, Albuquerque, NM 87131, USA}
\affiliation{Center for Biomedical Engineering, University of New Mexico, Albuquerque, NM 87131, USA}\author{Darko Stefanovic}
\affiliation{Department of Computer Science, University of New Mexico, Albuquerque, NM 87131, USA}
\affiliation{Center for Biomedical Engineering, University of New Mexico, Albuquerque, NM 87131, USA}
\affiliation{Nanoscience and Microsystems Engineering, University of New Mexico, Albuquerque, NM 87131, USA}

\begin{abstract}
We study the motion of random walkers with residence time bias between first and subsequent visits to a site, as a model for synthetic molecular walkers composed of coupled DNAzyme legs known as molecular spiders. The mechanism of the transient superdiffusion has been explained via the emergence of a boundary between the new and the previously visited sites, and the tendency of the multi-legged spider to cling to this boundary, provided residence time for a first visit to a site is longer than for subsequent visits. Using both kinetic Monte Carlo simulation and an analytical approach, we model a system that consists of single-legged walkers, each on its own one-dimensional track, connected by a “leash”, i.e., a kinematic constraint that no two spiders can be more than a certain distance apart. Even though a single one-legged walker does not at all exhibit directional, superdiffusive motion, we find that a team of one-legged walkers on parallel tracks, connected by a flexible tether, does enjoy a superdiffusive transient. Furthermore, the one-legged walker teams exhibit a greater expected number of steps per boundary period and are able to diffuse more quickly through the product sea than two-legged walkers, which leads to longer periods of superdiffusion.

\end{abstract}

\maketitle


\section{Introduction}

Cargo transport is ubiquitous in biological systems, and is often carried out by molecular walkers that are able to perform very specific actions despite the chaotic nature of the environment in which they act. Cyclic patterns of movement by biological molecular motors are typically the result of a catalyzed conformational change that converts chemical energy into directional movement~\cite{Schliwa:2003, Vale_2003, Ait-Haddou:2003, Krendel:2005, Geeves:2016}. These mechanisms are highly complex and still not fully understood and therefore pose a problem when we try to design and synthesize novel molecular walkers in the laboratory. Simpler walkers that do not rely on complex conformations can be constructed from DNA, which move stochastically and can be biased in a given direction through rational design of their environment and the track on which they move. The track can be a 2D surface from which DNA strands protrude, and the walker can be one or more DNA legs that move from site to site via branch migration. Some models use additional DNA strands in solution to block and unblock sites adjacent to the walker to achieve directional movement~\cite{Omabegho:2009,Muscat:2012}. Thubagere et~al.\ employed a diffusive walker that simply wanders about its track to pick up and deliver cargo to marked locations~\cite{Thubagere:2017}. 

We are interested in designs that employ restriction endonucleases~\cite{Bath:2005, Wickham:2011, Wickham:2012, Yang:2016, Ji:2017} or DNAzymes~\cite{Tian:2005,Lund:2010,Cha:2014,Cha:2015,Cai:2016,Yang:2018} to permanently modify sites via cleavage, which results in increased residence time during cleavage at previously unvisited sites versus already cleaved sites. With the proper design, a walker may preferentially explore a region of unvisited sites rather than an adjacent region of previously visited sites. When multiple legs are coupled to each other, the increased dwell time at new sites allows the other legs to either find an uncleaved site or to exclude other legs from moving toward the region of previously visited sites. Hence, these systems are affected greatly by the shape of the track and the geometric constraints of the walker(s). Inspired by Rank et al. \cite{Rank:2013}, we model the effect of the number of walker legs (1 or 2) and the length of a tether connecting multiple walkers on their own independent 1D tracks as tunable parameters that can be exploited to maximize the propensity for superdiffusive motion. We hope that this work will provide understanding to further geometrical optimizations of stochastic walkers in nanotechnology applications.

We build on previous models that are based on a class of synthetic molecular walkers known as molecular spiders~\cite{Antal:2007,Rank:2013,Semenov:2013_2,Semenov:2011,Semenov:2013}. A spider is composed of an inert chemical body and one or more flexible single-stranded DNAzyme legs. The DNAzyme is composed of two recognition arms separated by a catalytic core. Once the recognition arms bind to their complements, the catalytic core cleaves the bound strand at its single RNA base facing the catalytic core. Rates of association to and dissociation from recognition sites of the DNAzyme are a function of environmental factors such as ion concentration, choice of ion, and pH~\cite{Santoro:1998}. These factors influence the kinetics of the cleaving mechanism and may be optimized for maximum processive motion~\cite{Cha:2015}. Experimental data suggest the ratio of dwell time at uncleaved versus cleaved sites could be in the range $(0.05,0.1)$~\cite{Pei:2006}. The difference in residence times for legs at visited and unvisited sites has been studied extensively as it relates to first passage properties and directional movement in various simulated environments~\cite{Semenov:2013,Rank:2013,Olah:2013,Semenov:2011,Antal:2007}. Notable potential applications of the molecular spider include cargo transport or release of DNA or other product that occurs as a result of the enzymatic action~\cite{ECAL2013}. In both cases, sustained and consistent movement is a desirable quality, which we show can be tuned based on the geometry of the spider. Our model abstracts away all chemical kinetics and their factors and replaces them with a single parameter, the hopping rate $r$, where legs hop from previously cleaved sites at $r=1$ and dwell longer at new sites, hopping at rate $r<1$. Therefore the results apply not only to different types of molecular walkers, but any stochastic process that involves an object hopping to adjacent states with asymmetry in residence time between new and previously visited states.

Antal and Krapivsky~\cite{Antal:2007} analyze the simplest case of a single spider on a 1D track. They show that two-legged spiders exhibit transient superdiffusive motion while one-legged spiders do not. This is because the rear leg of the spider is likely to prevent a front leg that has just completed a cleavage to move in the negative direction away from new substrate sites, whereas a one-legged spider can move in either direction with equal probability. Semenov et al. \cite{Semenov:2011} further describe the motion of the spider as existing in one of two metastates, Boundary (B) and Diffusive (D).  The legs of the spider only move to adjacent sites, so for a 1D track this results in a growing contiguous region of cleaved sites, creating a distinct boundary between visited and unvisited regions. The B state is defined as any state in which at least one leg is on a substrate site, meaning it is at the boundary between visited and unvisited sites and is performing a cleavage. The D state is any state in which none of the spider legs are on a substrate site, and the spider is performing a random walk through the sea of previously visited sites. Ideally the spider would persist in the B state, cleaving substrates and moving away from the origin at a constant rate. Semenov et~al.\ show that a multi-legged spider is probabilistically biased toward unvisited sites while in the B state. When the B state ends, the spider enters the D state until it diffuses back to the boundary and begins another B state. Hence, there are two properties that we seek to optimize, the expected number of consecutive cleavages during a boundary period and the rate of diffusion during diffusive periods. 

We are motivated by the results of Rank et~al.~\cite{Rank:2013} on the effect of tethers between multiple two-legged spiders on separate parallel 1D tracks. They show that tethering multiple spiders on parallel tracks can increase the expected number of cleavages per spider during a boundary period, which varies as a function of the number of spiders per team and the length of the tether. They find that coupling spiders leads to a significant improvement in mean displacement and velocity, and reduction in variance compared with independent spiders. They provide a detailed analysis for the case of $r{\rightarrow}0$ that illustrates the benefit of tethered teams. They also provide simulations of motion for various $r$ values and show that an optimal tether length exists for non-zero values of $r$. However, this setting conflates the effects of multiple legs per spider with the effects of multiple spiders per team. In this work, we simplify the analysis of Rank et~al.~\cite{Rank:2013} by considering teams of single-legged spiders in order to elucidate the effect of the tether apart from the effect of multiple legs. Furthermore, we show that for non-zero $r$ values and a given tether length, teams of one-legged spiders are likely to remain in the B state longer than two-legged teams and that one-legged spider teams diffuse more quickly, the combined effect being a higher propensity for superdiffusive motion.

\begin{figure}[ht]
    \centering
    \includegraphics[width=0.95\columnwidth]{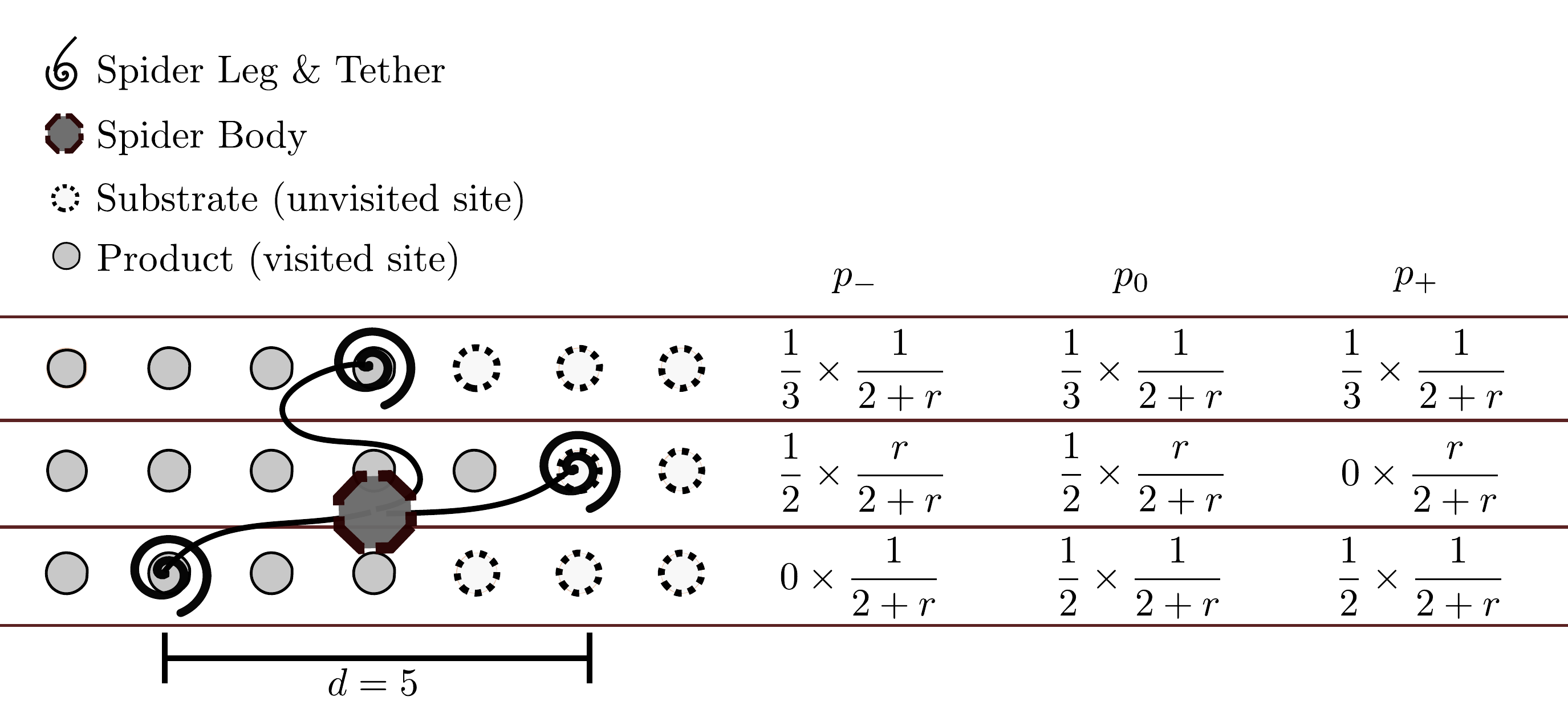}
    \caption{A team of three one-legged spiders ($w=3$) with tether length $d=5$ is depicted above. A spider leg will hop from its current position at rate $r=1$ if on a product site and $r<1$ if on substrate. A leg may move to any available position with uniform probability, where any position that does not violate the tether constraint is feasible. A leg moves at each step in our simulation. The probability for a leg to move left ($p_-$), right ($p_+$), or to its own location ($p_0$) is the probability that it will be the next leg to move (its rate divided by the total rate of all legs) multiplied by the probability of moving to the given location.}
    \label{fig:ModelDepiction}
\end{figure}

\section{Theoretical Analysis}
The position of a two-legged spider can be described fully by its center of mass. At integer values, its center of mass is at the unoccupied site between its legs; at half-integer values, its legs are at adjacent sites. A step is defined as the translation of a spider's center of mass by one. Antal and Krapivsky~\cite{Antal:2007} consider a two-legged spider that begins with its center of mass at the origin, where all sites in the positive direction are uncleaved (circles with hats), and all sites at or before the origin have been visited (circles without hats):

\begin{eqnarray}
\begin{array}{ccccccccc}  \arraycolsep.5pt {\ldots} & \circ & \circ & \bullet & \circ & \hat \bullet & \hat \circ & \hat \circ & {\ldots} \end{array} 
\label{AK_Init}
\end{eqnarray}

\noindent They derive the probability that a spider in configuration (\ref{AK_Init}) will take a whole step to the right or to the left as a function of the cleavage rate $r$, and show that as $r{\rightarrow}0$, the probability to step forward converges to the maximum value $p_+ = \frac{5}{8}$. Note that this definition implies that after the spider's forward leg has cleaved the substrate site, it can take any number of half-steps back and forth from its original position before completing a whole step in either direction. 

The special case of $r{\rightarrow}0$ would realistically not allow motion in the positive direction, but permits simplifications in the state space of the system. The case $r{\rightarrow}0$ is effectively the same as infinitely fast diffusion on visited sites. If two tethered spiders are both attached to substrate, each spider will cleave its substrate before the other with equal probability. Then, the spider will instantly diffuse to the next substrate position, given that it is within the range of the tether. If so, then due to the stochastic nature of the process, each spider will again cleave its substrate first with equal probability. Therefore, we can describe the state of a spider team solely by the difference in position between the leftmost substrates on each track. The team will move between states until all team members have cleaved every site within the range of the tether, and there is a single lagging spider attached to substrate.
This leads to the simplified state space for a team of two two-legged spiders, Equation 14 in Rank et~al.~\cite{Rank:2013}:

\begin{equation}
\arraycolsep.8pt \begin{array}{cccccccccc}
\bigl[ 0 \bigr] & \xrightleftharpoons[\frac 12]{1} & \bigl[ 1 \bigr] & \xrightleftharpoons[\frac 12]{\frac 12} & \dots & \xrightleftharpoons[\frac 12]{\frac 12} & \bigl[ d -1 \bigr] & \xrightleftharpoons[\Pi]{\frac 12} & \bigl[ d \bigr] \xrightarrow{1-\Pi} & \begin{minipage}{1.4cm} \textnormal{diffusive \\ period} \end{minipage} \, ,
\end{array} 
\label{eq_reaction_scheme_2Leg}
\end{equation}

\noindent and the analogous state space for a team of two one-legged spiders:

\begin{equation}
\arraycolsep.8pt \begin{array}{cccccccccc}
\bigl[ 0 \bigr] & \xrightleftharpoons[\frac 12]{1} & \bigl[ 1 \bigr] & \xrightleftharpoons[\frac 12]{\frac 12} & \dots & \xrightleftharpoons[\frac 12]{\frac 12} & \bigl[ d \bigr] & \xrightleftharpoons[\Pi]{\frac 12} & \bigl[ d + 1 \bigr] \xrightarrow{1-\Pi} & \begin{minipage}{1.4cm} \textnormal{diffusive \\ period} \end{minipage} \, ,
\end{array} 
\label{eq_reaction_scheme_1Leg}
\end{equation}

\noindent where the value in brackets is the absolute difference
between leftmost substrates, and $d$ is the length of the tether. Because $d$ is defined as the maximum distance between spider legs, we can see that the one-legged spider teams can reach one substrate further than two-legged spider teams, as the forward leg of a two-legged team is restricted by the rear leg of its teammate. Once the team reaches a potentially terminal state [$d$] (two-legged spiders) or [$d+1$] (one-legged spiders), it will continue the B state if the lagging spider steps to the right with the probability defined by Rank et al. as ${\Pi}$, the survival probability. They incorporate ${\Pi}$ into a derivation of the expected number of team steps $\langle S \rangle$ during the B state for $r{\rightarrow}0$ and prove the result through simulation, where a team step is defined by the average leftmost substrate position of all spiders increasing by one. This is Equation 17 in Rank et~al.~\cite{Rank:2013}: 

\begin{equation}
{\langle}S_{n=2}(r{\rightarrow}0){\rangle} = \dfrac{{\Pi}}{1-{\Pi}}+(d-1)\dfrac{{1}}{1-{\Pi}} 
\label{eq:Nsteps_2Leg}
\end{equation}

\noindent The analogous equation for two one-legged spiders is

\begin{equation}
{\langle}S_{n=1}(r{\rightarrow}0){\rangle} = \dfrac{d+{\Pi}}{1-{\Pi}} 
\label{eq:Nsteps_1Leg}
\end{equation}

Rank et~al.\ define ${\Pi}$ in the same way Antal and Krapivsky define $p_+$. Recall that this definition implies that after the spider's forward leg has cleaved the substrate site, it can take any number of half-steps back and forth from its original position before completing a whole step in either direction. This changes the meaning of the B state, since the probability for stepping forward includes a half-step in either direction from its original position after cleaving the substrate before a step in either direction has occurred. For example, from configuration (\ref{AK_Init}) after the right leg has completed its cleavage it may take a half-step away from its original position before the B state has terminated:

\begin{eqnarray}
\begin{array}{ccccccccc}  \arraycolsep.5pt {\ldots} & \circ & \circ & \bullet & \bullet & \circ & \hat \circ & \hat \circ & {\ldots} \end{array} 
\label{AK_Limbo}
\end{eqnarray}

\noindent The spider position one half-step further in the negative direction will occur with probability $1-{\Pi}$ thus terminating the B state. One and a half steps to the boundary in the positive direction will occur with probability ${\Pi}$ and preserve the B state. This means that after cleaving the substrate from configuration (\ref{AK_Init}), the spider can explore within $1\frac{1}{2}$ steps from the new boundary without terminating the B state, which will be referred to as ``limbo". This is different from the intuitive definition of the B state, that is, any state in which at least one leg is attached to substrate. 

To calculate the survival probability, we must look at the microstates composing the [$d$] state for a team of two two-legged spiders. This is Equation 15 in Rank et~al.~\cite{Rank:2013}:

\begin{eqnarray}\begin{aligned} & \arraycolsep.5pt  \circled{1} =\begin{array}{ccccc}  \arraycolsep.5pt \circ & \circ & \bullet & \bullet & \hat \circ \\\circ & \bullet & \hat \bullet & \hat \circ & \hat \circ \end{array} , \circled{2} =\begin{array}{ccccc} \circ & \bullet & \circ & \bullet & \hat \circ \\ \circ & \bullet & \hat \bullet & \hat \circ & \hat \circ \end{array} , \circled{3} =\begin{array}{ccccc} \circ & \bullet & \bullet & \circ & \hat \circ \\ \circ & \bullet & \hat \bullet & \hat \circ & \hat \circ \end{array}, \\
& \arraycolsep.5pt \circled{4} =\begin{array}{ccccc} \bullet & \circ & \bullet & \circ & \hat \circ \\ \circ & \bullet & \hat \bullet & \hat \circ & \hat \circ \end{array}, \circled{5} =\begin{array}{ccccc} \bullet & \bullet & \circ & \circ & \hat \circ \\ \circ & \bullet & \hat \bullet & \hat \circ & \hat \circ \end{array}, \circled{6} =\begin{array}{ccccc} \circ & \bullet & \bullet & \circ & \hat \circ \\ \bullet & \circ & \hat \bullet & \hat \circ & \hat \circ \end{array},  \\
& \arraycolsep.5pt \circled{7} =\begin{array}{ccccc} \bullet & \circ & \bullet & \circ & \hat \circ \\ \bullet & \circ & \hat \bullet & \hat \circ & \hat \circ \end{array}, \circled{8} =\begin{array}{ccccc} \bullet & \bullet & \circ & \circ & \hat \circ \\ \bullet & \circ & \hat \bullet & \hat \circ & \hat \circ \end{array} . \end{aligned} \label{2Leg_TermStateEqn}
\end{eqnarray} 

\noindent The analogous set of microstates for the [$d+1$] state of two one-legged spiders is:

\begin{eqnarray}\begin{aligned} & \arraycolsep.5pt  \circled{1} =\begin{array}{cccccc}  \arraycolsep.5pt \circ & \circ & \circ & \circ & \bullet & \hat \circ \\\circ & \circ & \hat \bullet & \hat  \circ & \hat \circ & \hat \circ \end{array} , \circled{2} =\begin{array}{cccccc}  \arraycolsep.5pt \circ & \circ & \circ & \bullet & \circ & \hat \circ \\\circ & \circ & \hat \bullet & \hat  \circ & \hat \circ & \hat \circ \end{array} ,\circled{3} =\begin{array}{cccccc}  \arraycolsep.5pt \circ & \circ & \bullet & \circ & \circ & \hat \circ \\\circ & \circ & \hat \bullet & \hat  \circ & \hat \circ & \hat \circ \end{array}  \\
& \arraycolsep.5pt \circled{4} =\begin{array}{cccccc}  \arraycolsep.5pt \circ & \bullet & \circ & \circ & \circ & \hat \circ \\\circ & \circ & \hat \bullet & \hat  \circ & \hat \circ & \hat \circ \end{array} , \circled{5} =\begin{array}{cccccc}  \arraycolsep.5pt \bullet & \circ & \circ & \circ & \circ & \hat \circ \\\circ & \circ & \hat \bullet & \hat  \circ & \hat \circ & \hat \circ \end{array} . \end{aligned} 
\label{1Leg_TermStateEqn}
\end{eqnarray} 

From these configurations the lagging spider (bottom track) will either step left or right, thus terminating or preserving the B state of the team. The possible configurations of the team are shown in matrix form in Figure~\ref{fig:MatrixOfStates}. In the $r{\rightarrow}0$ case, all of the configurations in (\ref{eq_reaction_scheme_2Leg}) and (\ref{eq_reaction_scheme_1Leg}) are equally probable at the time the final leg cleaves its substrate. From any one of these microstates we calculate the probability that the spider first takes a step to the right before taking a step to the left. The survival probability is then the average probability over all starting microstates (S) that the bottom spider steps right. Table \ref{TheorNSteps_Table} shows that, as expected, when the tether length becomes arbitrarily large, the survival probability for a team of two two-legged spiders converges to Antal and Krapivsky's $p_+=0.625$. For a team of two one-legged spiders, since the survival probability is constant and equal to $0.5$, equation (\ref{eq:Nsteps_1Leg}) simplifies to:

\begin{equation}
{\langle}S_{n=1,w=2}(r{\rightarrow}0){\rangle} = 2d+1
\label{eq:Nsteps_1Leg_simplified}
\end{equation}

\begin{figure}
    \subfloat[Two-Legged Team]{
    \centering
        \renewcommand{\arraystretch}{2}
        \setlength\tabcolsep{5pt}
        \begin{tabular}{|c||c|c|c|c|c|c|c|}
        \hline
        com   & -0.5 & 0 & 0.5 & 1 & 1.5 & 2 & 2.5    \\ \hline \hline
        0     & D & D & D &   &   &   &                 \\ \hline
        0.5   & L & L & L & L & L &   &                 \\ \hline
        1     &   &   & S & S & S &   &                 \\ \hline
        1.5   &   &   & S & S & S & S & S               \\ \hline
        2     &   &   &   &   & B & B & B               \\ \hline
        \end{tabular}}
        \hfill
    \subfloat[One-Legged Team]{
        \centering
        \renewcommand{\arraystretch}{2}
        \setlength\tabcolsep{6pt}
        \begin{tabular}{|c||c|c|c|c|c|}
        \hline
        com   & -1 & 0 & 1 & 2 & 3    \\ \hline \hline
        0     & D & D & D & D &       \\ \hline
        1     & S & S & S & S & S     \\ \hline
        2     &   & B & B & B & B     \\ \hline
        \end{tabular}}
        \caption{The entries of the matrices above are the possible states of spider teams with parameters $w=2$ and $d=2$. Row and column labels indicate the center of mass of the individual spiders, corresponding to the top and bottom spider states shown in equations (\ref{2Leg_TermStateEqn}) and (\ref{1Leg_TermStateEqn}). Labels (B) and (D) indicate the start of boundary and diffusive states. Label (S) indicates the possible starting locations for a spider entering limbo as shown in (\ref{eq_reaction_scheme_2Leg}) and (\ref{eq_reaction_scheme_1Leg}). Label (L) indicates a spider in the limbo state. The survival probability is the average probability that a spider starting in any state (S) will reach any state (B) before any state (D). For the one-legged team, we can see by symmetry that the survival probability is $\frac{1}{2}$.}
    \label{fig:MatrixOfStates}
\end{figure}

\begin{table}[ht]
\renewcommand{\arraystretch}{1.5}
\setlength\tabcolsep{4pt}
\begin{tabular}{|c||c|c|c|c|}
\hline
    & \multicolumn{2}{c|}{1 Leg}       & \multicolumn{2}{c|}{2 Legs}          \\ 
d   & $\Pi$  & $\langle${}S$\rangle{}$ & $\Pi$     & $\langle${}S$\rangle${} \\ \hline\hline
2   & 0.5 & 5                          & 0.6534 & 4.7705                     \\
4   & 0.5 & 9                          & 0.6291 & 9.7846                     \\
8   & 0.5 & 17                         & 0.6267 & 20.433                     \\
16  & 0.5 & 33                         & 0.6258 & 41.759                     \\
32  & 0.5 & 65                         & 0.6254 & 84.423                     \\
64  & 0.5 & 129                        & 0.6252 & 169.75                     \\
128 & 0.5 & 257                        & 0.6251 & 340.42                     \\ \hline
\end{tabular}    
\caption{Analytically derived survival probabilities ${\Pi}$ and expected number of steps $\langle S \rangle$ as a function of leash length $d$ for one and two-legged spider teams of two spiders ($w=2$). $\langle S \rangle$ increases proportional to $d$, as shown in equations (\ref{eq:Nsteps_2Leg}) and (\ref{eq:Nsteps_1Leg_simplified}).}
\label{TheorNSteps_Table}
\end{table}

\begin{figure}[ht]
    \centering
    \includegraphics[width=0.95\columnwidth]{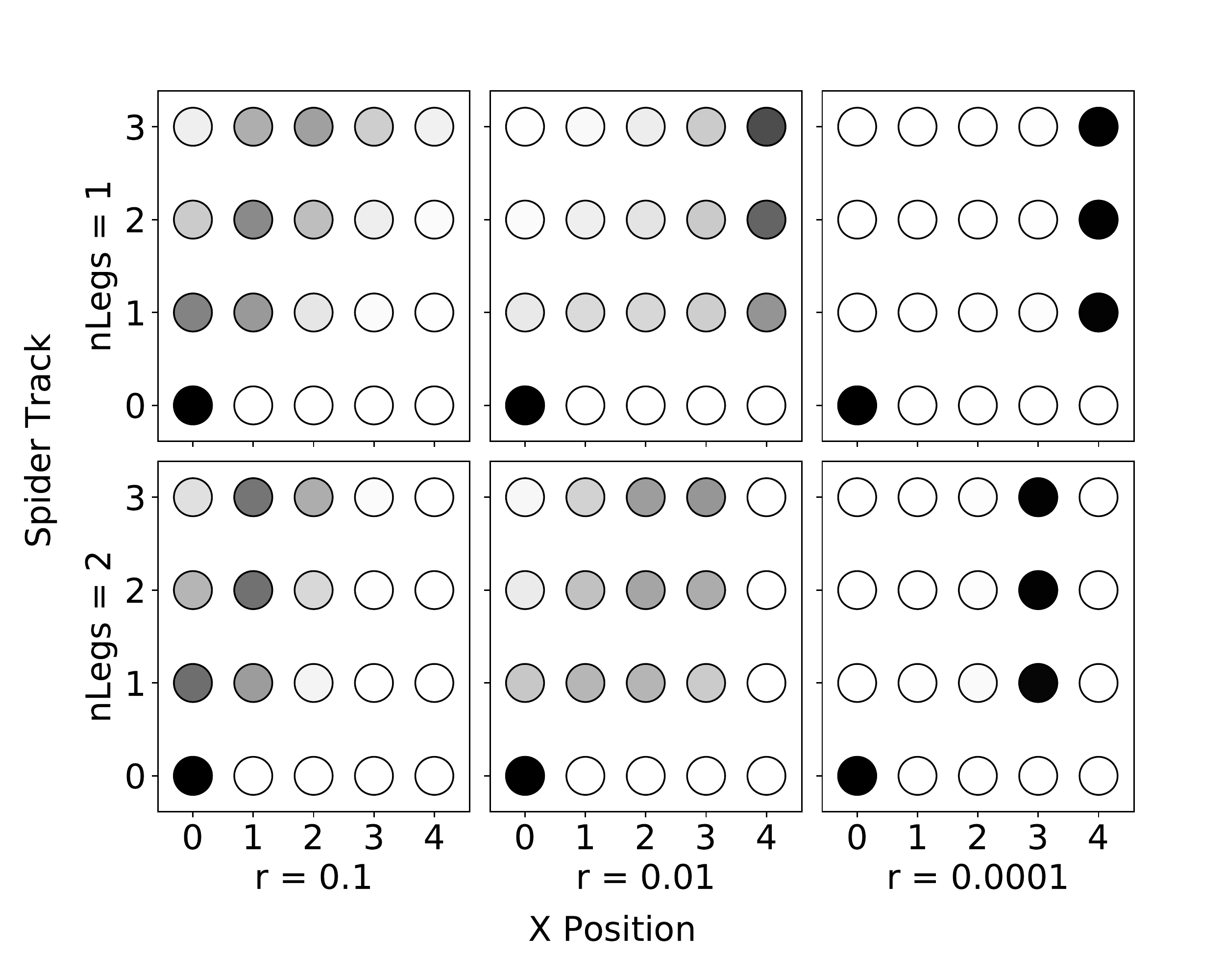}
    \caption{The distribution of substrates was recorded at the time the boundary period terminated and normalized to the origin. The tracks (y-axis) are ranked by distance of the substrate position nearest the origin, and shaded according to the probability density. Shown is the result for parameters $w=4$ and $d=4$.}
    \label{fig:TermStates}
\end{figure}

\section{Simulation}

Sites are initialized asymmetrically with only substrate sites in the positive direction from the origin and only product sites otherwise. A spider leg can detach from its current site and then instantly move to any site within its constraints (including the same site). Spider legs cannot overtake one another, thus the order of front and rear legs is maintained at all times. The tether imposes the constraint that the (rear) leg of any spider cannot be more than the specified distance from the (front) leg of any other spider, as if all spiders are attached to a common body by their own tether, as shown in Figure~\ref{fig:ModelDepiction}. Pei et~al.~\cite{Pei:2006} provide experimental data for DNAzyme spiders from which we derived the ratio of dwell times at visited to unvisited sites in the range $(0.05,0.1)$. We ue the value $r=0.1$ to elucidate the effect of the various geometries at realistic DNAzyme $r$ values, as it is furthest from $r{\rightarrow}0$, which maximizes $\langle S \rangle$. 

\section{Results}

\subsection{Expected Team Steps During a Boundary Period}

\begin{figure}[ht]
\subfloat[One-Legged Team]{
    \centering
        \centering
        \includegraphics[width=0.95\columnwidth]{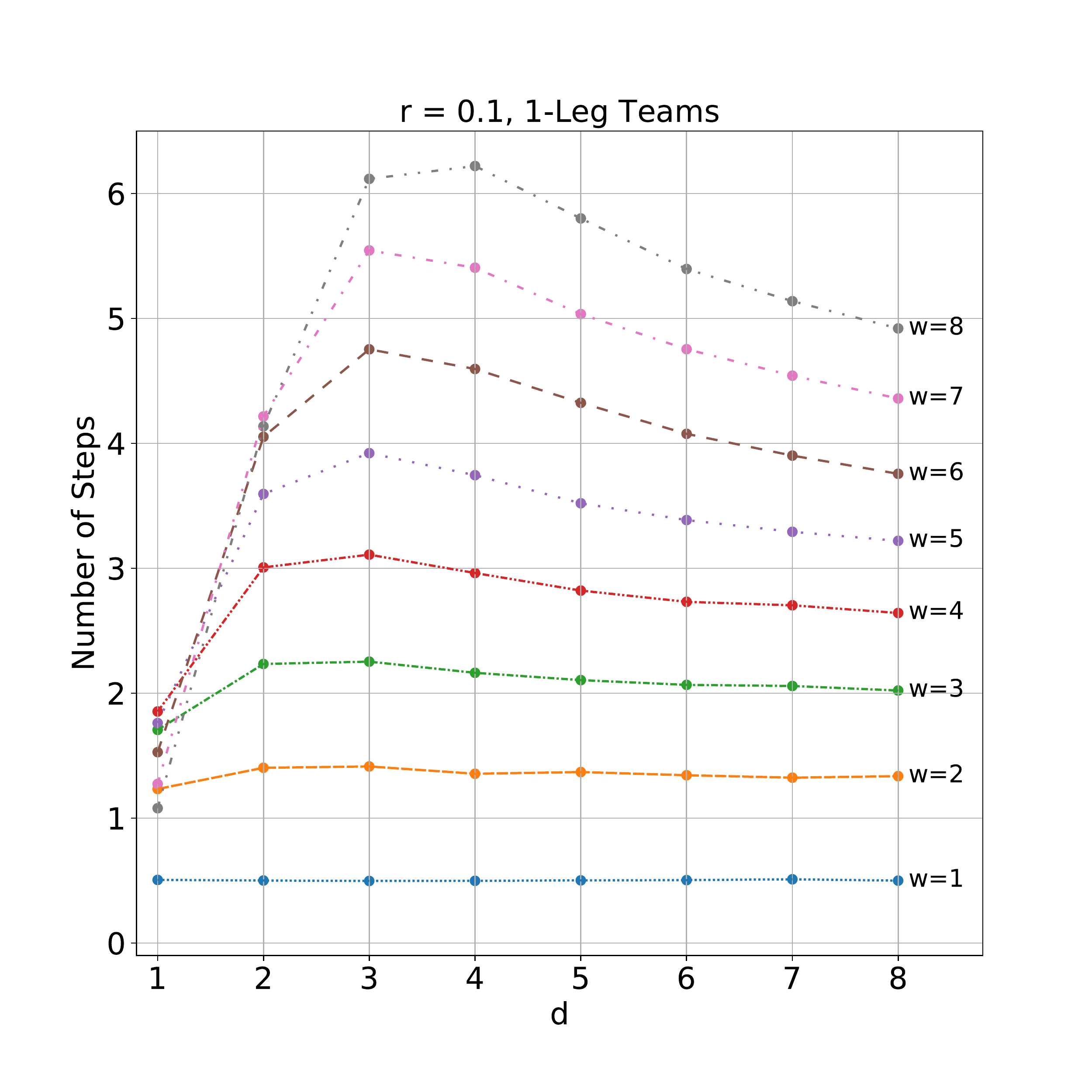}}
    \hfill
    \subfloat[Two-Legged Team]{
        \centering
        \includegraphics[width=0.95\columnwidth]{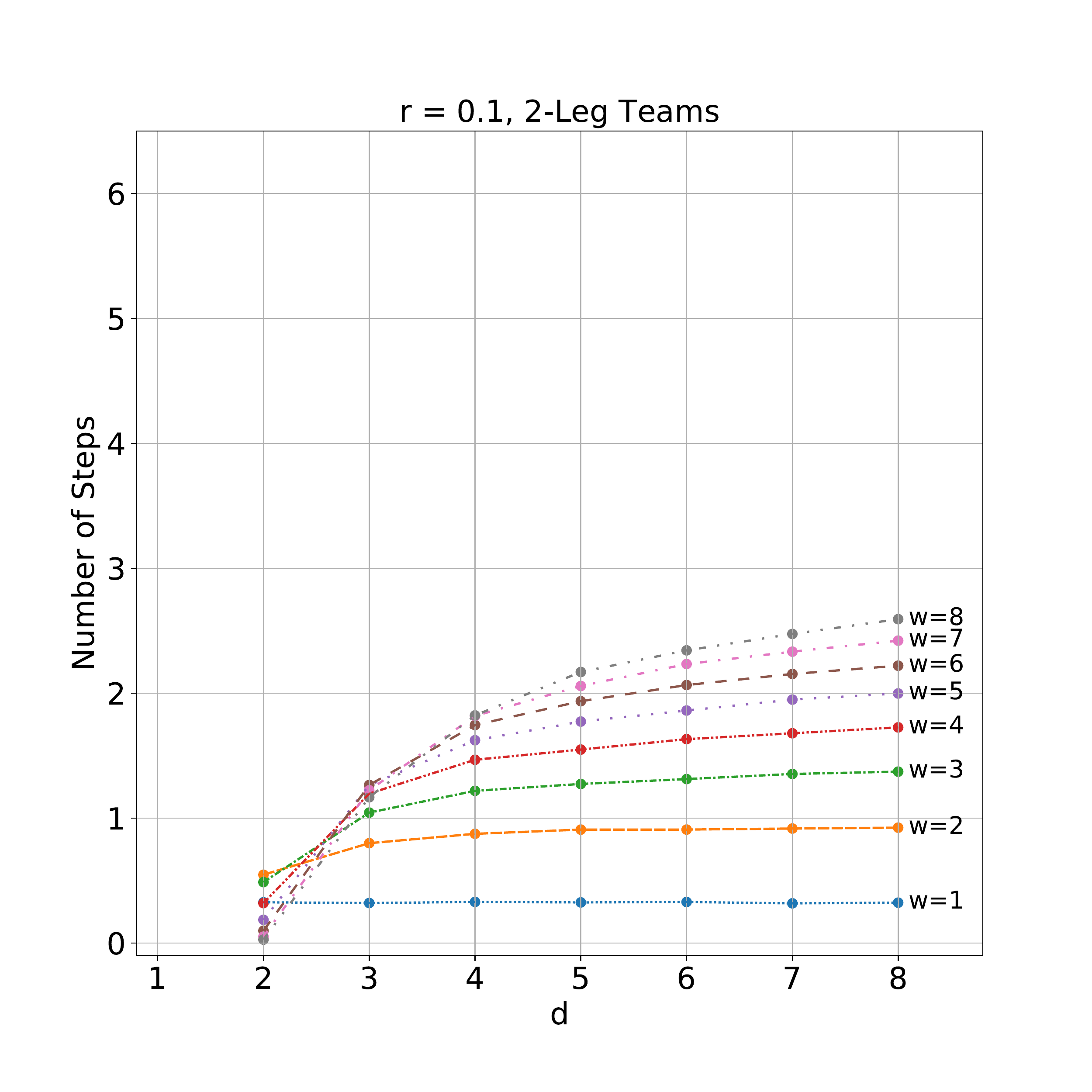}}
    \caption{Simulated expected number of steps per boundary period. Note that any movement is prohibited for 2-legged teams with a tether of length 1. Spiders are initialized at the origin with all legs on substrate.}
    \label{fig:NSteps_Realistic}
\end{figure}

\begin{figure}[htp]
\includegraphics[width=0.95\columnwidth]{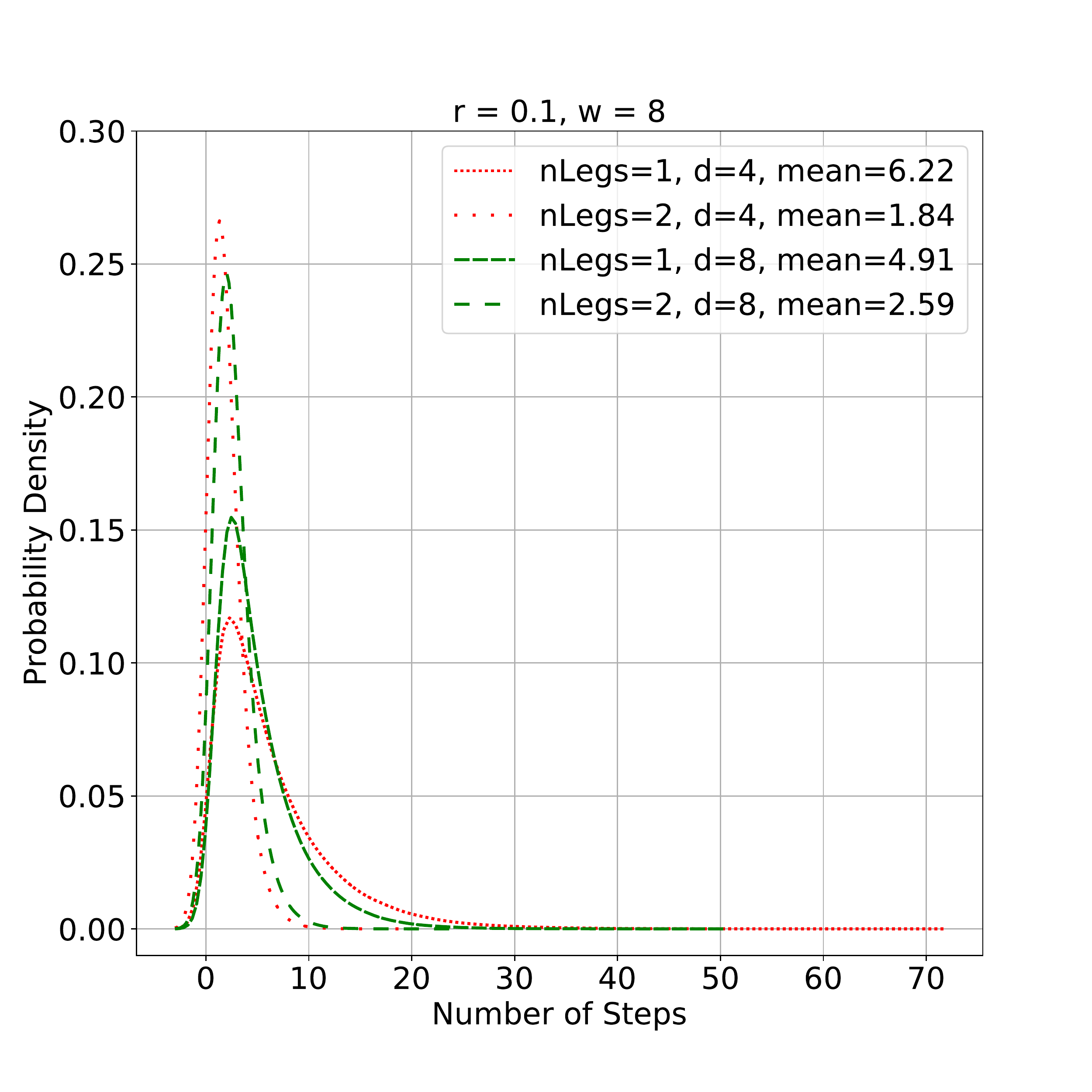}
\caption{Probability density for number of steps per boundary period. Curves were generated using the kernel density estimate in seaborn~\cite{seaborn}.}
\label{NSteps_Distribution}
\end{figure}

For these simulations we use the intuitive definition of the B state, viz., any state in which at least one spider on the team is performing a cleavage. We follow Rank et~al.\ in defining a team step to occur when the average leftmost substrate position over all tracks increases by one. We look at the best-case scenario for maximum expected steps where all spiders are initialized on substrate and the absolute difference between all substrate locations is zero. Table~\ref{TheorNSteps_Table} shows that the two-legged spider teams have higher survival probability and a greater expected number of steps during a B period for the $r{\rightarrow}0$ case; however, the one-legged teams perform much better at realistic $r$ values as seen in Figure~\ref{fig:NSteps_Realistic}. The distribution of steps during $2\times10^6$ boundary periods for a subset of team configurations is shown in Figure~\ref{NSteps_Distribution}. A notable difference between the non-zero $r$ and $r{\rightarrow}0$ cases is that a spider which has detached from its own boundary is not guaranteed to find its boundary before another spider detaches from its boundary. The spider team may exit into a diffusive period from any of the states shown in (\ref{eq_reaction_scheme_2Leg}) and (\ref{eq_reaction_scheme_1Leg}) for non-zero $r$. We recorded the relative substrate positions at the time each boundary period terminated, and show that the spider team approaches the behavior described in the theoretical analysis of $r{\rightarrow}0$ as $r$ decreases (Figure~\ref{fig:TermStates}). 

\subsection{Superdiffusive Motion}

For $r=0.1$, we record the position of spider teams defined as the mean position of all legs and compare the results for one and two-legged spider teams. The mean displacement for an unbiased random walk in 1D is $\langle x \rangle=0$. The biased random walk will have a positive mean displacement as a function of time. The distribution of displacements for $10^4$ simulations is shown for the same subset of parameters in Figure~\ref{NSteps_Distribution}. 

\begin{figure}[ht]
    \centering
    \subfloat[Mean Square Displacement vs. Time. Thick black guide lines represent slopes corresponding to ballistic (${\alpha}$=2) and diffusive (${\alpha}$=1) motion.]{
        \centering
        \includegraphics[width=0.95\columnwidth]{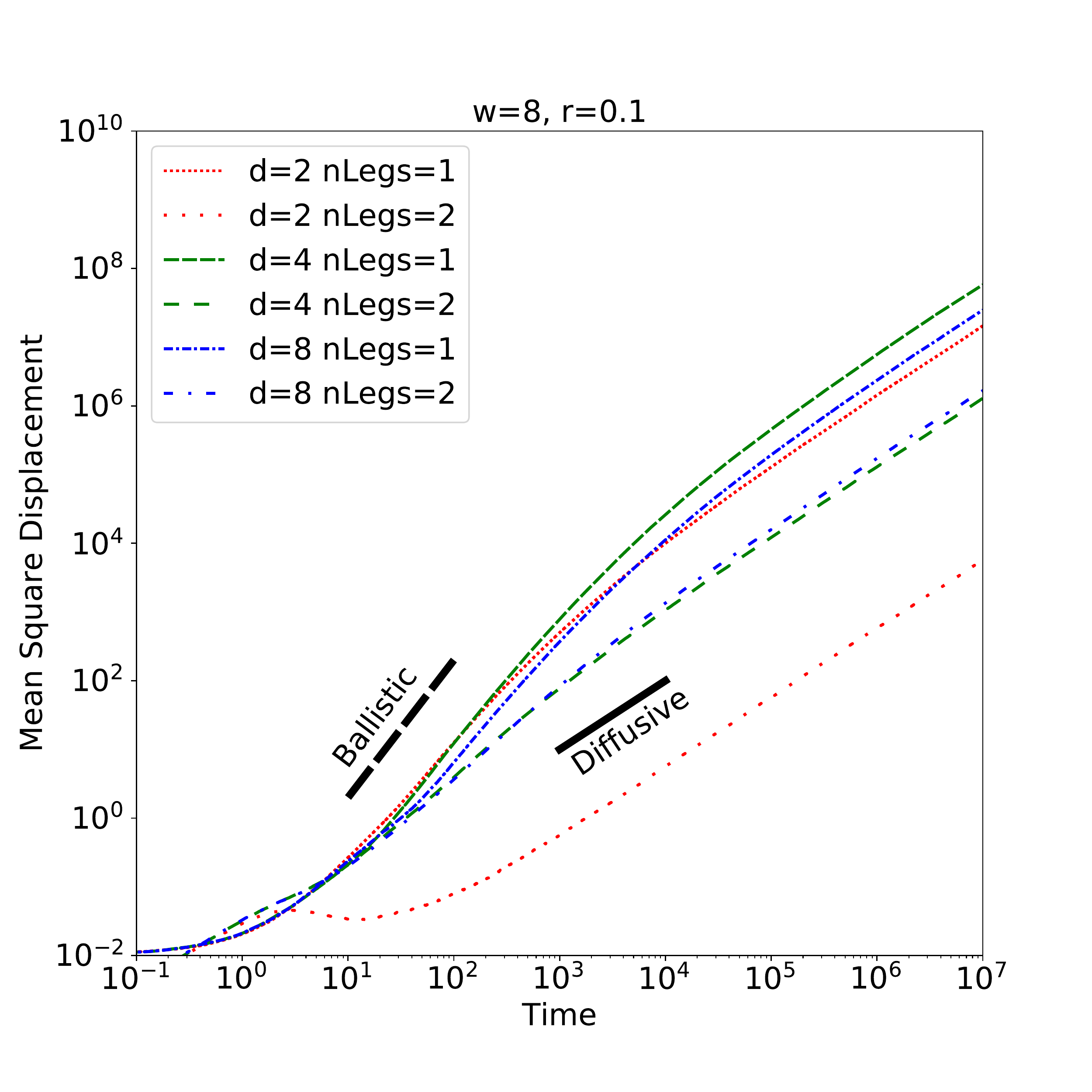}}
    \hfill
    \subfloat[Finite difference approximation of $\alpha$ from msd data, smoothed using Savitzky-Golay filtering \cite{Press:2002}.]{
        \centering
        \includegraphics[width=0.95\columnwidth]{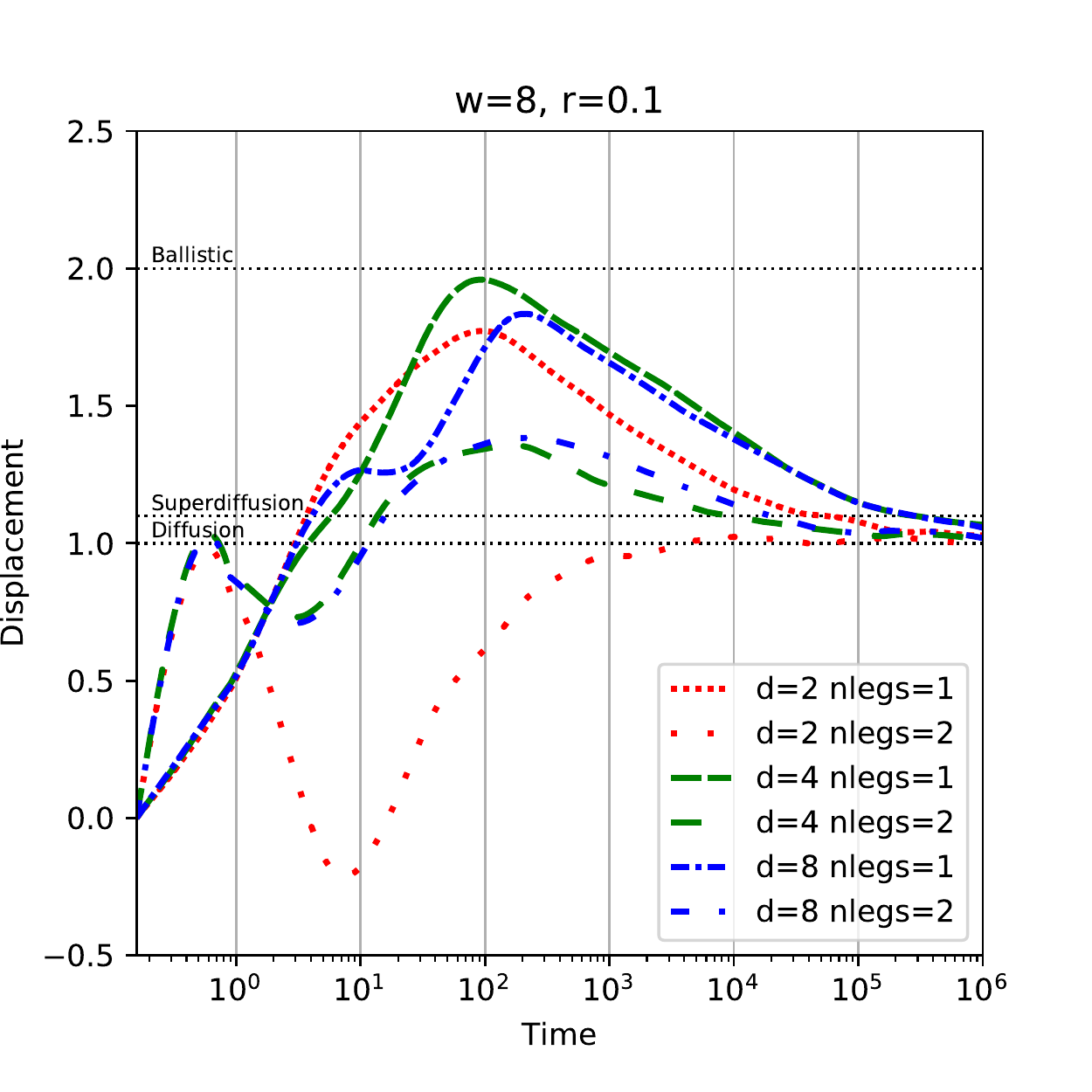}}
    \caption{Mean Square Displacement and finite difference approximation of $\alpha$ for $10^4$ traces with parameters $w=8$, $r=0.1$, and $d=2,4,8$.}
    \label{fig:msd}
\end{figure}

The mean square displacement (MSD) in one-dimensional space is defined as
\begin{equation}
\langle x^2\rangle = 2Dt^\alpha,
\end{equation}
where $D$ is the diffusion constant, and is affected by the geometry of the spider and number of spiders per team. The parameter $\alpha$ is equal to $1$ for simple diffusion and equal to $2$ for constant directional velocity. We use the finite difference approximation of our MSD data to calculate the parameter $\alpha$, as shown in Figure~\ref{fig:msd},
\begin{equation}
\label{alpha}
\alpha=\frac{d(log \langle x^2(t)\rangle )}{d(log(t))}.
\end{equation}

\noindent  We chose the arbitrary threshold $\alpha=1.1$ to denote significant superdiffusive motion. Mean values for $\alpha$ and MSD are shown in Figure \ref{fig:msd}.

\begin{figure}[ht]
    \centering
    \subfloat[]{
        \centering
        \includegraphics[width=0.95\columnwidth]{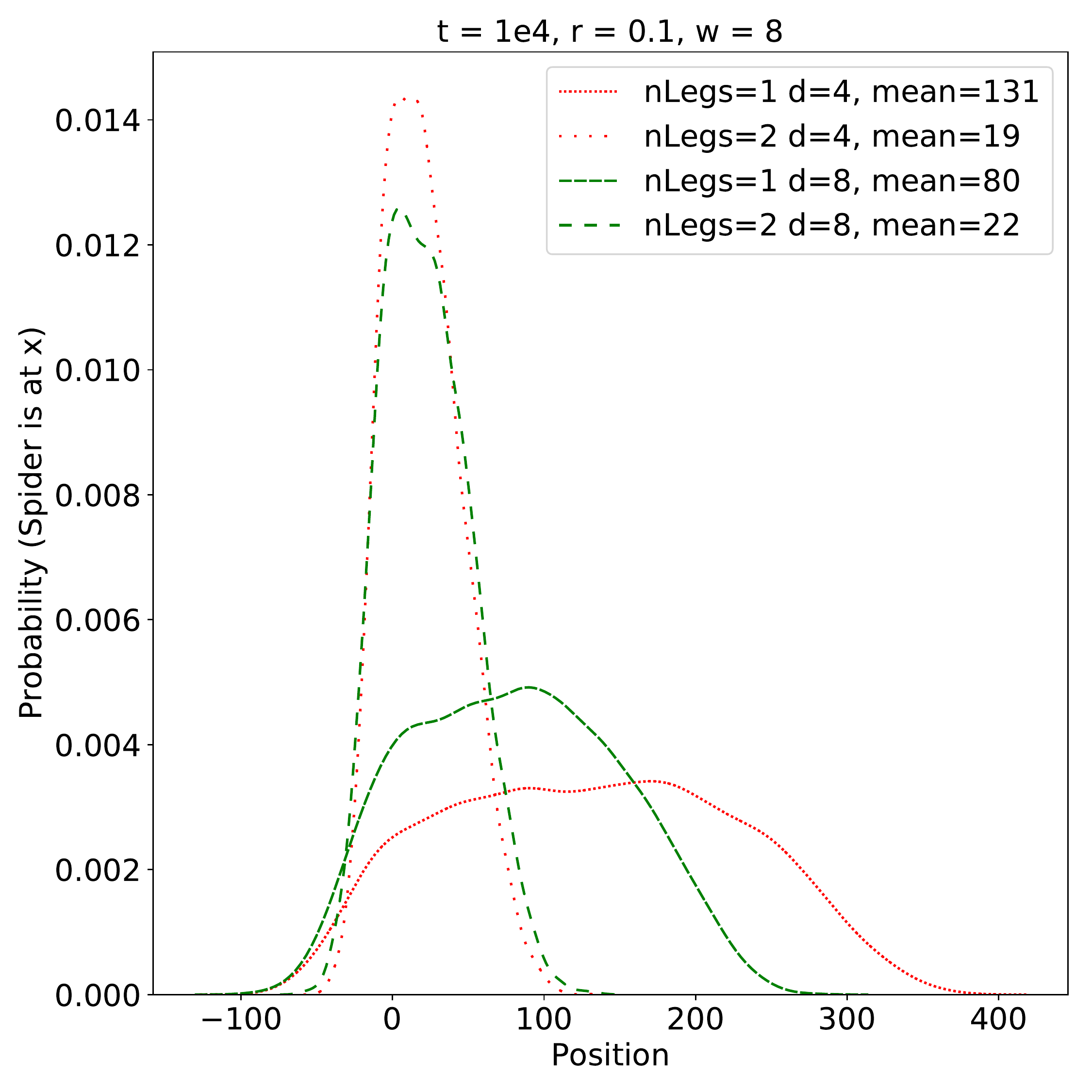}}
    \hfill
    \subfloat[]{
        \centering
        \includegraphics[width=0.95\columnwidth]{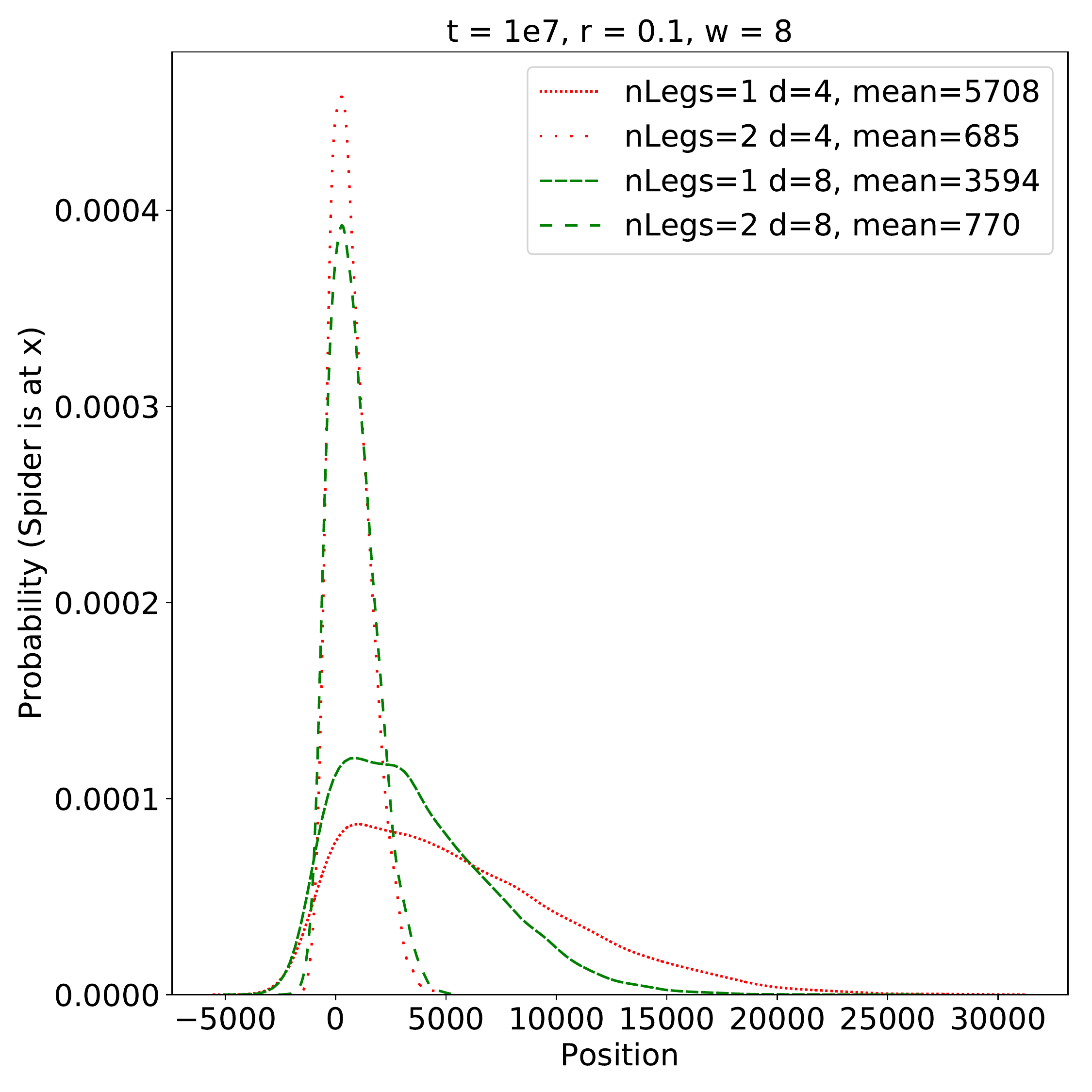}}
    \caption{Probability Density of spider positions at times $10^4$ (a) and $10^7$ (b) for teams of 8 spiders at rate $r=0.1$. Curves were generated using the   kernel density estimate in seaborn~\cite{seaborn}.}
    \label{fig:hist}
\end{figure}

\section{Discussion}

The system in Figure~\ref{fig:ModelDepiction} depicts a team of one-legged spiders tethered to each other on independent 1D tracks, where a leg visits and cleaves a new site (substrate) with rate $r<1$, and visits already cleaved sites (product) at rate $1$. The system was analyzed for the theoretical case, $r{\rightarrow}0$, and simulated using Kinetic Monte Carlo simulations for $r=0.1$, which represents a realistic rate for the DNAzyme spider~\cite{Pei:2006}. 

We recapitulate the $r{\rightarrow}0$ analysis of Rank et~al.~\cite{Rank:2013} for two-legged spiders tethered to one another on independent 1D tracks. In this limit, any spider effectively finds the next substrate site on its track instantly after completing a cleavage, so long as the next substrate is within the range of the tether. They define a boundary period that begins when any spider attaches to a substrate site for the first time and ends when all spiders have taken one step away from the boundary. The simplified state space for one and two-legged walker teams in the $r{\rightarrow}0$ limit is compared in equations (\ref{eq_reaction_scheme_2Leg}) and (\ref{eq_reaction_scheme_1Leg}). The expected number of steps during a boundary period $\langle S \rangle$ follows from these state spaces, see equations (\ref{eq:Nsteps_2Leg}) and (\ref{eq:Nsteps_1Leg}). These equations are dependent on the survival probability $\Pi$, defined as the probability of stepping right for the case when all spiders have cleaved every substrate within the range of the tether, and there is a single lagging spider performing a cleavage. This probability is derived for teams of two spiders as a function of the tether length $d$. For arbitrarily large $d$, the result agrees with the analysis of Antal and Krapivsky~\cite{Antal:2007} for single spiders in 1D. Indeed, two-legged spiders have a higher survival probability that leads to a greater $\langle S \rangle$ during a boundary period in the limit $r{\rightarrow}0$ as shown in Table~\ref{TheorNSteps_Table}. We find that this is not the case for realistic values of $r$.

The state space for teams of two walkers shown in equations (\ref{eq_reaction_scheme_2Leg}) and (\ref{eq_reaction_scheme_1Leg}) do not apply to realistic $r$ values. Because cleavages occur on a finite timescale, a spider that has cleaved its substrate is not guaranteed to diffuse to its next substrate site before another team member moves away from its boundary. This is shown in Figure~\ref{fig:TermStates}. The desired effect of the tether is to keep a team member that has detached from its boundary in close enough proximity such that it can diffuse to its boundary again before the last substrate-bound team member has completed cleavage. Rank et al. show that there exists an optimal tether length for non-zero values of $r$ that is dependent on the number of spiders on the team, the cleavage rate, and the tether length. Intuitively, the optimal length is long enough such that the lagging spiders do not impede the forward motion of their teammates, but short enough such that a spider that has detached from its boundary has enough time to diffuse to its boundary. Because one-legged spiders diffuse faster than two-legged spiders, this effectively increases the probability that a spider detached from its boundary will diffuse back to its boundary before the last spider has detached from substrate. Semenov et~al.~\cite{Semenov:2011} provide the theoretical diffusion constant $D=0.25$ along with the observed $D=0.247 \pm 0.010$ for a single two-legged spider, compared to the Einstein relation for Brownian motion in 1D, $D={l^2}/{2\tau}$, which simplifies to $D=0.5$ for step size $l=1$ and expected time between moves defined by the rate $r=1$ for one-legged spiders. Hence, the faster diffusion of one-legged spiders benefits these teams by increasing the probability that a spider will diffuse to the next substrate site before its team detaches from the boundary. Figure \ref{fig:NSteps_Realistic} shows a significant increase in $\langle S \rangle$ for one-legged versus two-legged spider teams. Furthermore, Figure~\ref{NSteps_Distribution} shows that in addition to a higher mean number of steps, the tails of these distributions extend further with higher probability for one-legged teams. 

Increasing the number of spiders per team results in greater $\langle S \rangle$, and also shifts the optimal tether length in the positive direction, as shown in Figure~\ref{fig:NSteps_Realistic}. This is because a greater number of spiders decreases the probability that all will simultaneously detach from the boundary, and in turn increases the time window for a detached spider to find a substrate site. With regard to mean square displacement, this effect is opposed by a decreasing team diffusion rate. Rank et~al.~\cite{Rank:2013} show that the diffusion constant for a team of two two-legged walkers with leash length $d=2$ is roughly $D=0.04$, and $D=0.02$ for a team of three. With enough spiders for a fixed tether length, the team as a whole would not be able to progress, as a lagging spider would consistently thwart the forward motion of its teammates. The optimal number of team members warrants further investigation.

Semenov et~al.~\cite{Semenov:2011} describe the motion of spiders as existing in one of two metastates, the boundary state B and the diffusive state D. The higher rate of diffusion for one-legged spiders not only benefits $\langle S \rangle$, but also contributes to reducing the time spent in diffusive periods by allowing the team to more quickly find the boundary. These effects together result in greater distances traveled by one-legged versus two-legged teams over long time periods as well as longer periods of more superdiffusive motion, as seen in Figures~\ref{fig:hist}~and~\ref{fig:msd}. 

Notably, our description of the system as a team of one-legged spiders can equivalently be seen as a single spider with multiple legs, each of which is constrained to its own independent track. While this study only focuses on the case of independent one-dimensional tracks, we provide insight into the properties of stochastic spiders that promote superdiffusive motion, namely the role of the rate of diffusion and constraints that affect the expected number of steps per boundary period. We hope that these insights may inspire further investigation into other geometrical configurations of spiders and tracks, which might lead to the discovery of a system with effectively infinite periods of directional ballistic motion. 

\section{Acknowledgments}
We thank Erwin Frey and Matthias Rank for help understanding the details of their simulation. This material is based upon work supported by the National Science Foundation under grant 1422840.

\bibliography{sources}
\end{document}